\documentclass[12pt,epsf]{article}
\usepackage{graphicx}
\usepackage{epsfig}
\begin{document}
%%%%%%%%%%%%%%%%%%%%%%%%%%%%%%%%%%%%%%%%%%%
 
\def\a{\alpha}
\def\b{\beta}
\def\c{\varepsilon}
\def\d{\delta}
\def\e{\epsilon}
\def\f{\phi}
\def\g{\gamma}
\def\h{\theta}
\def\k{\kappa}
\def\l{\lambda}
\def\m{\mu}
\def\n{\nu}
\def\p{\psi}
\def\q{\partial}
\def\r{\rho}
\def\s{\sigma}
\def\t{\tau}
\def\u{\upsilon}
\def\v{\varphi}
\def\w{\omega}
\def\x{\xi}
\def\y{\eta}
\def\z{\zeta}
\def\D{\Delta}
\def\G{\Gamma}
\def\H{\Theta}
\def\L{\Lambda}
\def\F{\Phi}
\def\P{\Psi}
\def\S{\Sigma}

\def\o{\over}
\def\beq{\begin{eqnarray}}
\def\eeq{\end{eqnarray}}
\newcommand{\gsim}{ \mathop{}_{\textstyle \sim}^{\textstyle >} }
\newcommand{\lsim}{ \mathop{}_{\textstyle \sim}^{\textstyle <} }
\newcommand{\vev}[1]{ \left\langle {#1} \right\rangle }
\newcommand{\bra}[1]{ \langle {#1} | }
\newcommand{\ket}[1]{ | {#1} \rangle }
\newcommand{\EV}{ {\rm eV} }
\newcommand{\KEV}{ {\rm keV} }
\newcommand{\MEV}{ {\rm MeV} }
\newcommand{\GEV}{ {\rm GeV} }
\newcommand{\TEV}{ {\rm TeV} }
\def\diag{\mathop{\rm diag}\nolimits}
\def\Spin{\mathop{\rm Spin}}
\def\SO{\mathop{\rm SO}}
\def\O{\mathop{\rm O}}
\def\SU{\mathop{\rm SU}}
\def\U{\mathop{\rm U}}
\def\Sp{\mathop{\rm Sp}}
\def\SL{\mathop{\rm SL}}
\def\tr{\mathop{\rm tr}}

\def\IJMP{Int.~J.~Mod.~Phys. }
\def\MPL{Mod.~Phys.~Lett. }
\def\NP{Nucl.~Phys. }
\def\PL{Phys.~Lett. }
\def\PR{Phys.~Rev. }
\def\PRL{Phys.~Rev.~Lett. }
\def\PTP{Prog.~Theor.~Phys. }
\def\ZP{Z.~Phys. }

%added by FT
\newcommand{\bear}{\begin{array}}  
\newcommand {\eear}{\end{array}}
\newcommand{\la}{\left\langle}  
\newcommand{\ra}{\right\rangle}
\newcommand{\non}{\nonumber}  
\newcommand{\ds}{\displaystyle}
\newcommand{\red}{\textcolor{red}}
\def\ubl{U(1)$_{\rm B-L}$}
\def\REF#1{(\ref{#1})}
\def\lrf#1#2{ \left(\frac{#1}{#2}\right)}
\def\lrfp#1#2#3{ \left(\frac{#1}{#2} \right)^{#3}}
\def\OG#1{ {\cal O}(#1){\rm\,GeV}}

%%%%%%%%%%%%%%%%%%%%%%%%%%%%%%%%%%%%%%%%%%%%%%%%%%%%%%%%%%%%%%%%%%%%

\baselineskip 0.7cm

\begin{titlepage}

\begin{flushright}
UT-09-29\\
IPMU 09-0159
\end{flushright}

\vskip 1.35cm
\begin{center}
{\large \bf
Low Scale Direct Gauge Mediation 
with \\ Perturbatively Stable Vacuum
}
\vskip 1.2cm
Ryosuke Sato and Kazuya Yonekura 
\vskip 0.4cm

{\it   Department of Physics, University of Tokyo,\\
    Tokyo 113-0033, Japan\\ 
 Institute for the Physics and Mathematics of the Universe (IPMU), 
University of Tokyo,\\ Chiba 277-8568, Japan}

\vskip 1.5cm

\abstract{In a class of direct gauge mediation with a perturbatively stable SUSY breaking vacuum, 
gaugino masses vanish at the leading order of SUSY breaking $F$-term.
We study the allowed parameter space of the gauge mediation models. By imposing a Tevatron bound
on the lightest chargino mass $m_{{\tilde \chi}^\pm_1} \gsim 270~\GEV$ and a warm dark-matter mass bound on the 
light gravitino mass $m_{3/2}\lsim 16~\EV$, we find that 
almost all the parameter space is excluded. Near future experiments may completely exclude, or possibly discover, the scenario.}
\end{center}
\end{titlepage}

\setcounter{page}{2}

%%%%%%%%%%%%%%%%%%%%%%%%%%%%%%%%
\section{Introduction}
\label{sec:1}
%%%%%%%%%%%%%%%%%%%%%%%%%%%%%%%%
Low scale direct gauge mediation~\cite{Giudice:1998bp} is very attractive. 
It can achieve a very light gravitino mass $m_{3/2} \lsim 16~\EV$, satisfying the constraint from cosmology~\cite{Viel:2005qj}.
SUSY breaking vacuum can be sufficiently stable, even if not absolutely stable.
Furthermore, low scale gauge mediation can be tested in future experiment at the LHC~\cite{Hinchliffe:1998ys}.

An explicit model of direct mediation with dynamical SUSY breaking was first constructed in Ref.~\cite{Izawa:1997gs} (see also Ref.~\cite{Ibe:2005xc})
using the Izawa-Yanagida-Intriligator-Thomas (IYIT) model~\cite{Izawa:1996pk} as a 
SUSY breaking sector. Then, after the discovery of the Intriligator-Seiberg-Shih (ISS) metastable SUSY breaking model~\cite{Intriligator:2006dd}, 
many works have been done~\cite{Kitano:2006xg} to construct a direct mediation model by gauging a flavor symmetry of the ISS model. 
However, in those works it was observed~\cite{Izawa:1997gs,Ibe:2005xc,Kitano:2006xg} that the Minimal Supersymmetric Standard Model (MSSM) gaugino masses are suppressed.
That is, the gaugino masses vanish at the leading order of a SUSY breaking $F$ term, in an expansion in $F/m^2$ where $m$ is a messenger mass scale. 
Later, the vanishing of the leading term in the gaugino masses is shown~\cite{Komargodski:2009jf} in general set up.
If a model can be described at low energies by some weakly coupled O'Raifeartaigh type effective field theory and the SUSY breaking vacuum of the model is the 
lowest energy state of the effective theory, the leading term of the gaugino masses vanish. 

To obtain the gaugino masses comparable to sfermion masses, one has to choose one of the following two possibilities. The first possibility is  
to give up the stability of the vacuum and go to a higher metastable vacuum~\footnote{
Note that this metastability is the one which is present even in the low energy effective field theory at perturbative level. 
The original ISS model is considered to be ``stable'' in this sense. (Non-perturbative dynamical SUSY restoration is not taken into account.) }
\cite{Giveon:2009yu}. Such a vacuum becomes 
more unstable as we lower the gravitino mass (see Ref.~\cite{Hisano:2007gb} for a detailed study in the case of a so-called minimal gauge mediation).
The other possibility is to have a messenger mass scale $m$ and a SUSY breaking scale $\sqrt{F}$ to be comparable, i.e., $m \sim \sqrt{F}$.
Then, higher order terms in the gaugino masses (of order ${\cal O}(F^3/m^5)$) are not so small compared 
with the sfermion masses (of order ${\cal O}(F/m)$). 
The condition $m \sim \sqrt{F}$ is also required
to achieve the light gravitino mass.

In this paper, we study calculable low scale direct gauge mediation in which a SUSY breaking vacuum is stable at least at perturbative level.
Then the gaugino masses are suppressed due to the vanishing of the leading order term.
There are two important constraints on such a gauge mediation scenario; the constraint on the lightest chargino mass $m_{{\tilde \chi}^\pm_1} \gsim 270~\GEV$
from the CDF collaboration at the Tevatron collider~\cite{Aaltonen:2009tp} (and also the recent work Ref.~\cite{Meade:2009qv}),
and the constraint on the light gravitino mass $m_{3/2} \lsim 16~\EV$ from 
the warm dark matter bound~\cite{Viel:2005qj}.\footnote{
The gravitino mass bound is not applicable for a heavy enough gravitino mass and/or a low enough reheating temperature.
But note that such a heavy gravitino mass (i.e. large SUSY breaking scale $F$) means heavy (i.e. well beyond ${\cal O}(1~\TEV)$) sfermion masses 
due to the $F/m^2$ suppression of the gaugino masses. Furthermore, a low reheating temperature is inconsistent with many Baryogenesis scenarios, 
so we do not consider such a case in this paper.}
These constraints give lower and upper bounds on the SUSY breaking scale, respectively, 
so it is not obvious whether there is any parameter space consistent with the bounds.
The aim of this paper is to study the allowed parameter space. 
We find that there is almost no parameter space consistent with the above bounds.
There are some points in the parameter space that are near the margin of the bound,
and at such points the gluino is quite light, $m_{\tilde{g}} \lsim 800~\GEV$. Future experiments at the Tevatron (with enough integrated luminosity) 
or at the LHC may decide whether the scenario studied in
this paper is relevant to nature or not.

This paper is organized as follows. In section \ref{sec:2}, we classify low scale direct gauge mediation models with a stable vacuum. 
Then in section \ref{sec:3} we calculate the maximum value of the lightest chargino mass $m_{{\tilde \chi}^\pm_1}$ and the gluino mass $m_{\tilde{g}}$
under the condition that Yukawa interaction of the messenger sector is perturbative up to the GUT scale. 
We compare the result with the experimental bound.
In section \ref{sec:iss} we consider a direct mediation in the ISS model,
as an exceptional model in which the condition of section \ref{sec:3} is not satisfied. The last section is devoted to conclusions and discussion.

%%%%%%%%%%%%%%%%%%%%%%%%%%%%%%%%
\section{Classification of direct mediation models}
\label{sec:2}
%%%%%%%%%%%%%%%%%%%%%%%%%%%%%%%%

In this section, we generalize the model discussed in Refs.~\cite{Izawa:1997gs,Ibe:2005xc}.
A direct gauge mediation model using the Intriligator-Seiberg-Shih (ISS) model will be discussed in section~\ref{sec:iss}.
We take the IYIT model~\cite{Izawa:1996pk} as a SUSY breaking sector for concreteness, but our result does not depend on this choice
and any type of generalized O'Raifeartaigh models may work as well.
At low energies, the effective superpotential is given by
\begin{eqnarray}
W \simeq \Lambda^2 Z.
\end{eqnarray}
Here, $Z$ is a singlet chiral superfield, and $\Lambda$ the SUSY breaking scale, respectively.
We introduce $N_F$ flavors of messenger multiplets which transform as {\bf 5} and ${\bf {\bar 5}}$ under the GUT gauge group $SU(5)_{{\rm GUT}}$.
We will write the messenger quark multiplets as $\Psi_{d,i},{\tilde \Psi_{d,i}}\ (i=1,\cdots,N_F)$
 and the messenger lepton multiplets as $\Psi_{l,i},{\tilde \Psi_{l,i}}$.

We assume that the superpotential of the messenger sector is given as follows :
\begin{eqnarray}
W &=& \Lambda^2 Z + \sum_{\chi = d,l} \sum_{i,j} M_{\chi,ij}(Z) {\tilde \Psi}_{\chi,i} \Psi_{\chi,j}, \\
&&\left( M_{\chi,ij}(Z)=m_{\chi,ij} + k_{\chi,ij} Z \right). \nonumber
\end{eqnarray}
We require that the model satisfies the following conditions.
\begin{enumerate}
\item The SUSY breaking vacuum with $\la \Psi \ra = {\tilde {\langle\Psi\rangle}} = 0$ is stable.
\item The model has an R-symmetry.
\item A $D$ term of $U(1)_{\rm Y}$ is not generated by messenger loops at 1-loop level.
\item There is no CP violation in the model.
\item The Standard Model~(SM) gauge couplings and the messenger sector Yukawa couplings 
are perturbatively small up to the GUT scale, $2\times 10^{16}\ \GEV$.
\end{enumerate}

The first condition in particular requires that ${\rm det} (m+kZ) = {\rm det}m$ \cite{Komargodski:2009jf}.
In order for the messengers to be massive, we should impose that ${\rm det} (m+kZ) \neq 0$.
Then, ${\rm det} m \neq 0$.

The second condition is imposed to maintain the SUSY breaking~\cite{Nelson:1993nf} (see also Ref.~\cite{Shih:2007av}). 
It is known that dynamical SUSY breaking models often have an R-symmetry.
The IYIT model has an R-symmetry, and we assume that the messenger sector also respects that symmetry. Without the R-symmetry,
introduction of ``generic'' Yukawa couplings between $Z$ and the messengers restores SUSY~\cite{Nelson:1993nf}. 
However, note that this second condition does not completely ensure the first condition imposed above, 
because of the existence of a runaway behavior. See below.

As to the third condition, if a non zero $U(1)_{\rm Y}$ $D$ term is generated at 1-loop,
the squared masses of the sfermions become negative.
We introduce a messenger parity to avoid such a danger \cite{Dimopoulos:1996ig}~(see also Ref.~\cite{Buican:2008ws}). 
The messenger parity is a symmetry that transforms the messenger chiral superfields $\Psi_i$, ${\tilde \Psi}_i$ and 
the $U(1)_{\rm Y}$ vector superfield $V$ as follows
\footnote{Of course $SU(3)_C$ and $SU(2)_L$ should also be transformed in an appropriate way.}
\begin{eqnarray}
\Psi_i \to \sum_j U_{ij} {\tilde \Psi}_j,\ {\tilde \Psi}_i \to \sum_j {\tilde U}_{ij} \Psi_j,\ V \to -V,
\end{eqnarray}
where $U$ and ${\tilde U}$ are some unitary matrices.
Owing to this symmetry, the dangerous $D$ term is not generated by messenger loops.

In order for the fourth condition to be satisfied, we assume that $m_\chi$ and $k_\chi$ are real for the time being.
However, we will show that $m_\chi$ and $k_\chi$ can be taken real without loss of generality in the model which can make the gaugino masses maximum.

The last condition requires that $N_F \leq 4$,
otherwise the perturbative gauge coupling unification is lost \cite{Jones:2008ib}.

We investigate $M(Z)$ satisfying those conditions.
First, we simplify the messenger parity transformation by changing the basis of the messengers and redefining the transformation.
We define $\P'_i \equiv \sum_j X_{ij} \P_j$ and ${\tilde \P}'_i \equiv \sum_j (XU)_{ij} {\tilde \P}_j$, 
where $X$ is some unitary matrix.
Then, the messenger parity transformation is given by
\begin{eqnarray}
\P'_i \to {\tilde\P}'_i,\ {\tilde\P}'_i \to \sum_j (XU{\tilde U}X^\dagger)_{ij} \P'_j,\ V \to -V.
\end{eqnarray}
Since $U{\tilde U}$ is a unitary matrix, we can take $X$ to diagonalize $U\tilde U$.
Then the messenger parity transformation is given by
\begin{eqnarray}
\P'_i \to {\tilde \P}'_i,\ {\tilde \P}'_i \to e^{i\b_i} \P'_i,\ V\to -V.
\end{eqnarray}
When we operate the messenger parity transformation twice,
$\P'$ and ${\tilde\P}'$ are transformed as,
\begin{eqnarray}
\P'_i \to e^{i\b_i} \P'_i,\ {\tilde \P}'_i \to e^{i\b_i} {\tilde \P}'_i.
\end{eqnarray}
This transformation is also a symmetry transformation of the model. 
Then, in this basis,
\begin{eqnarray}
m_{ij} = e^{i(\b_i+\b_j)} m_{ij},\ 
k_{ij} = e^{i(\b_i+\b_j)} k_{ij}.
\end{eqnarray}

We classify $\b_i$ as follows :
\begin{eqnarray}
\b_i = \pi && (i=1,\cdots,N_\pi), \\
-\pi < \b_i <\pi && (i=N_\pi+1,\cdots, N_F).
\end{eqnarray}
For $1\leq i\leq N_\pi$ and $N_{\pi}+1 \leq j\leq N_F$,
$0<\b_i+\b_j<2\pi$, then $m_{ij}=m_{ji}=k_{ij}=k_{ji}=0$.
Therefore $m$ and $k$ are block-diagonal matrices which are given by
\begin{eqnarray}
m=\left(
\begin{array}{cc}
m_\pi&\\
&m_0\\
\end{array}
\right),\ 
k=\left(
\begin{array}{cc}
k_\pi&\\
&k_0\\
\end{array}
\right).
\end{eqnarray}
For $N_\pi+1 \leq i,j \leq N_F$, $-2\pi<\b_i+\b_j<2\pi$,
then, $\b_i+\b_j = 0$ or $m_{ij} = k_{ij}= 0$.
Now, we show that the following transformation is a symmetry transformation of the model.
\begin{eqnarray}
\P_i'\to Y_i \P_i',~{\tilde\P}'_i \to Y_i {\tilde\P}'_i,
\end{eqnarray}
where
\begin{eqnarray}
Y_i=\left\{
\begin{array}{ll}
1 & i=1,\cdots,N_\pi\\
e^{-i\beta_i/2} & i=N_\pi+1,\cdots,N_F
\end{array}
\right.
.
\end{eqnarray}
The proof is as follows. Under this transformation,
\begin{eqnarray}
\sum_{i,j} m_{ij} {\tilde\P}'_i \P'_j \to \sum_{i,j} Y_i m_{ij} Y_j {\tilde\P}'_i \P'_j.
\end{eqnarray}
Because $\b_i+\b_j = 0$ for $i,j=N_\pi+1,\cdots ,N_F$ for which $m_{ij} \neq 0$,
\begin{eqnarray}
YmY = \left(
\begin{array}{cc}
m_\pi&\\
&m_{ij} e^{-i(\b_i+\b_j)/2}
\end{array}
\right)
 = \left(
\begin{array}{cc}
m_\pi&\\
&m_{ij}
\end{array}
\right).
\end{eqnarray}
Therefore, the mass term is invariant. It is easy to check that the Yukawa interaction term is also invariant.

By using the $Y$ transformation, we define a new messenger parity transformation.
After operating the original messenger parity transformation, we operate the $Y$ transformation.
Under this transformation, $\P'$, ${\tilde \P}'$ and $V$ are transformed as 
\begin{eqnarray}
\P'_i \to {\tilde\P}'_i,&{\tilde \P_i}'\to -\P'_i &(i=1,\cdots,N_\pi ),\nonumber \\
\P'_i \to e^{-i\b_i/2} {\tilde\P}'_i,&{\tilde \P_i}'\to e^{i\b_i/2} \P_i' &(i=N_\pi+1,\cdots,N_F ),\nonumber \\
V \to -V.
\end{eqnarray}
From now on, we call this new transformation as the messenger parity transformation.
By using appropriate phase rotation, we can take a new basis $\P_i$ and ${\tilde\P}_i$ in which the messenger parity transformation is given as follows :
\begin{eqnarray}
\P_i \to {\tilde\P}_i,&{\tilde \P_i}\to -\P_i &(i=1,\cdots,N_\pi ),\nonumber \\
\P_i \to {\tilde\P}_i,&{\tilde \P_i}\to \P_i &(i=N_\pi+1,\cdots,N_F ),\nonumber \\
V \to -V.
\end{eqnarray}
Because $m$ and $k$ are block-diagonal, we consider the cases $N_\pi = 0$ and $N_\pi=N_F$ separately.
Then, we have $U=1$ and $\tilde{U}=\pm 1$.
In more general cases, the messenger sector is a direct sum of these two cases.

Let us next consider about R-symmetry.
We assumed that there is an R-symmetry.
By using this R-symmetry and the messenger parity,
we can define the following symmetry transformation,
\begin{eqnarray}
\P \to \exp\left(i\a U{\tilde R}U^{-1} \right)\P = e^{i\a \tilde{R}}\P,\\ 
{\tilde \P} \to \exp\left(i\a {\tilde U}R{\tilde U}^{-1}\right) {\tilde \P} = e^{i\a R} {\tilde \P},
\end{eqnarray}
where $R$ is the generator of the R-symmetry on $\P$ and ${\tilde R}$ is on ${\tilde \P}$. We have used $U=1$ and $\tilde{U}=\pm 1$.
Under this transformation, the superpotential $W$ has R-charge 2. Therefore this transformation also generates an R-symmetry.
Then, we can define a new R-symmetry transformation as the sum of the original and the above R-symmetries,
\begin{eqnarray}
\P \to e^{i\a(R+\tilde{R})/2} \P,~
{\tilde \P} \to e^{i\a({\tilde R}+R)/2} {\tilde \P}.
\end{eqnarray}
It is obvious that this new R-symmetry commutes with the messenger parity. 
Therefore, we can assign R-charges as $R(\P_i) = R({\tilde \P}_i)$ by simultaneously diagonalizing these symmetries.
From now on, we call this new R-symmetry as the R-symmetry of the model.

We assign R-charges as $R(\Psi_1) \leq R(\Psi_2) \leq \cdots \leq R(\Psi_{N_F})$. 
We now show that $R(\Psi_{i}) + R(\Psi_{N_F-i+1}) = 2$.
If $R(\Psi_{i}) + R({\tilde \Psi}_{N_F-i+1}) < 2$, $\P_1,\cdots,\P_{i}$ cannot have mass terms with ${\tilde \P}_1, \cdots ,{\tilde \P}_{N_F-i+1}$.
Then the mass matrix is of the form,
\beq
m=\left(
\bear{cc}
0_{i,N_F-i+1} & * \\
{*}& *  
\eear
\right),
\eeq
where $0_{i,N_F-i+1}$ is a $i \times (N_F-i+1)$ matrix with all components equal to 0. This matrix has $\det m=0$.
One can also show that if $R(\Psi_{i}) + R({\tilde \Psi}_{N_F-i+1}) > 2$, then $\det m=0$.
This contradicts with the stability condition discussed above.
Thus, we must have $R(\Psi_{i}) + R({\tilde \Psi}_{N_F-i+1}) =2$.

We write $R(\Psi_1)$ as $a$. If $W$ includes an interaction term $kZ \Psi_1 {\tilde \Psi}_{N_F-i+1}$, we have $R({\tilde \Psi}_{N_F-i+1}) = -a$, and $R(\Psi_i) = a+2$.
Then if $W$ includes an interaction term $kZ \Psi_i {\tilde \Psi}_{N_F-j+1}$, $R({\tilde \Psi}_{N_F-j+1})=-a-2$ and $R(\Psi_{j}) = a+4$.
In this way, there are $\Psi$ which have R-charges $a,a+2,\cdots$, 
and $\tilde{\P}$ which have R-charges $-a+2,-a,\cdots$. 
We call them as $a$-group.
If there are some $\Psi$ which are not included in $a$-group, we write the smallest R-charge of them as $b$.
Such $\Psi$ make up $b$-group.
Therefore, all $\Psi$ are classified into the groups, $a$-group, $b$-group, $c$-group and so on.
The components of $M_\chi (Z)$ are nonzero only within each group. Note that if we take one group, $R(\P_{i+1})-R(\P_i)=0~{\rm or}~2$.

Some groups have a messenger parity by itself. If not, a group (assume it to be $a$-group) has an ``anti-group'' (which is $(-a-2n)$-group, 
where $(-a-2n)$ is the smallest R-charge of $\tilde{\P}$ contained in $a$-group) and they make up a pair
to exhibit the messenger parity symmetry.
However, we can combine these two groups into a single group by defining a new R-symmetry.
Since the components of $M_\chi(Z)$ are nonzero only within each group,
%
%there is a $U(1)$ symmetry which assigns charge $x$ to $\P$ in $a$-group and $\tilde \P$ in $(-a-2n)$-group,
%$-x$ to $\tilde \P$ in $a$-group and $\P$ in $(-a-2n)$-group,
%and 0 to other fields, where $x$ is an arbitrary number.
%Then, by using this $U(1)$ symmetry with $x=-(a+n)$, 
%we can define a new R-symmetry which assigns charge $-n,~-n+2,~\cdots$ to $\P$
%
there is a $U(1)$ symmetry which assigns charge $x$ to $\P$ in $a$-group, 
$-x$ to $\tilde \P$ in $a$-group,
$y$ to $\P$ in $(-a-2n)$-group,
$-y$ to $\tilde \P$ in $(-a-2n)$-group,
and 0 to other fields, where $x$ and $y$ are arbitrary numbers.
Then, by using this $U(1)$ symmetry with $x=-y=-(a+n)$, 
we can define a new R-symmetry which assigns charge $-n,~-n+2,~\cdots,~n+2$ to $\P$
and $n+2,~n,~\cdots,~-n$ to $\tilde \P$ in both of the groups.
Thus, we can combine $a$-group and $(-a+2n)$-group into a single $(-n)$-group, which has the messenger parity by itself.
Therefore, we can assume that the messenger sector is a direct sum of groups with each group having the messenger parity by itself. 
Note that we can maintain $R(\P_{i+1})-R(\P_i)=0~{\rm or}~2$ in the combined group after appropriately rearranging $\P$ such that $R(\P_i)\leq R(\P_{i+1})$.

In summary, we have obtained $R(\Psi_i)=R(\tilde{\Psi}_i)$, $ R(\Psi_i) + R(\Psi_{N_F-i+1}) = 2$,
and $R(\P_{i+1})-R(\P_i)  = 0~{\rm or}~2$ in one group. 
If we determine an R-charge assignment of the messengers following the above conditions, we can get the form of $M_\chi(Z)$.

First, we consider the case $U={\tilde U} = 1$, and there is only one group. As noted above, more general cases can be expressed as a direct sum.
The messenger parity transformation is given by $\Psi_i \leftrightarrow {\tilde \Psi}_i$,
and thus $M_\chi(Z) = M_\chi^T(Z)$.

\begin{itemize}

\item $N_F=1$ model :

In this model, $M_\chi (Z) = m$, that is, the messenger has no Yukawa interactions.

\item $N_F=2$ model : 

%$R(\Psi_2) = R(\Psi_1) + 2$.  ($R(\P_1)=0$.)
$(R(\P_1),R(\P_2))=(a,a+2),~~a=0$.
\begin{eqnarray}
\left(
\begin{array}{cc}
kZ & m \\
m & \\
\end{array}
\right).
\end{eqnarray}

\item $N_F=3$ model : 

%$R(\Psi_2) = R(\Psi_1) + 2$, $R(\Psi_3) = R(\Psi_1) + 4$.  ($R(\P_1)=-1$.)
$(R(\P_1),R(\P_2),R(\P_3))=(a,a+2,a+4),~~a=-1$.
\begin{eqnarray}
\left(
\begin{array}{ccc}
&kZ & m \\
kZ&m' & \\
m&&
\end{array}
\right).
\end{eqnarray}

\item $N_F=4$ model A :

%$R(\Psi_2) = R(\Psi_1) + 2$, $R(\Psi_3) = R(\Psi_1) + 4$, $R(\Psi_4) = R(\Psi_1) + 6$.  ($R(\P_1)=-2$.)
$(R(\P_1),R(\P_2),R(\P_3),R(\P_4))=(a,a+2,a+4,a+6),~~a=-2$.
\begin{eqnarray}
\left(
\begin{array}{cccc}
&&kZ & m \\
&k'Z&m' & \\
kZ&m'&& \\
m&&&
\end{array}
\right).
\end{eqnarray}

\item $N_F=4$ model B :

%$R(\Psi_2) = R(\Psi_1) + 2$, $R(\Psi_3) = R(\Psi_1) + 2$, $R(\Psi_4) = R(\Psi_1) + 4$.  ($R(\P_1)=-1$.)
$(R(\P_1),R(\P_2),R(\P_3),R(\P_4))=(a,a+2,a+2,a+4),~~a=-1$.
\begin{eqnarray}
\left(
\begin{array}{cccc}
&k_1Z&k_2Z & m_1 \\
k_1Z&m_3&m_2 & \\
k_2Z&m_2&m_4& \\
m_1&&&
\end{array}
\right).
\end{eqnarray}

\item $N_F=4$ model C :

%$R(\Psi_2) = R(\Psi_1)$, $R(\Psi_3) = R(\Psi_1) + 2$, $R(\Psi_4) = R(\Psi_1) + 2$.  ($R(\P_1)=0$.)
$(R(\P_1),R(\P_2),R(\P_3),R(\P_4))=(a,a,a+2,a+2),~~a=0$.
\begin{eqnarray}
\left(
\begin{array}{cccc}
k_2Z&k_1Z&m_3& m_1 \\
k_1Z&k_3Z&m_2 &m_4 \\
m_3&m_2&& \\
m_1&m_4&&
\end{array}
\right).
\end{eqnarray}

\end{itemize}
In fact, some of the above models have a runaway direction in the tree level potential.
See \ref{app-C} for details.
We must require that $km^{-1}k = 0$ for the model to have no runaway direction.
$N_F=3$ model has a runaway direction.
$N_F=4$ model A has a runaway direction unless $k'=0$ or $k=0$.
By unitary rotations, we can take $k_2=0$ in $N_F=4$ model B. Then this model has a runaway direction
unless $m_4=0$. $N_F=4$ model A is included in $N_F=4$ model B (if $k'=0$) or $N_F=2$ model (if $k=0$).
There is no runaway direction in $N_F=4$ model C. 

So far, we have considered the cases in which there is only one group.
We can consider a direct sum of two $N_F=2$ models, but this is included in $N_F=4$ model C.

Next, we consider the case $U=1,~{\tilde U} = -1$.
The messenger parity transformation is given by $\P_i \to {\tilde \P}_i$, ${\tilde \P}_i \to -\P_i$,
and thus $M_\chi(Z) = -M_\chi^T(Z)$.
In this case, $N_F$ is even, otherwise $\det M_\chi(Z) = 0$.

\begin{itemize}

\item $N_F=2$ model : 

%$R(\Psi_2) = R(\Psi_1) + 2$
$(R(\P_1),R(\P_2))=(a,a+2),~~a=0$.
\begin{eqnarray}
\left(
\begin{array}{cc}
 & m \\
-m & \\
\end{array}
\right).
\end{eqnarray}
In this model, the messengers have no Yukawa interactions.

\item $N_F=4$ model A' :

%$R(\Psi_2) = R(\Psi_1) + 2$, $R(\Psi_3) = R(\Psi_1) + 4$, $R(\Psi_4) = R(\Psi_1) + 6$, $R(\P_1)=-2$.
$(R(\P_1),R(\P_2),R(\P_3),R(\P_4))=(a,a+2,a+4,a+6),~~a=-2$.
\begin{eqnarray}
\left(
\begin{array}{cccc}
&&kZ & m \\
&&m' & \\
-kZ&-m'&& \\
-m&&&
\end{array}
\right).
\end{eqnarray}

\item $N_F=4$ model B' :

%$R(\Psi_2) = R(\Psi_1) + 2$, $R(\Psi_3) = R(\Psi_1) + 2$, $R(\Psi_4) = R(\Psi_1) + 4$, $R(\P_1)=-1$.
$(R(\P_1),R(\P_2),R(\P_3),R(\P_4))=(a,a+2,a+2,a+4),~~a=-1$.
\begin{eqnarray}
\left(
\begin{array}{cccc}
&k_1Z&k_2Z & m_1 \\
-k_1Z&&m_2 & \\
-k_2Z&-m_2&& \\
-m_1&&&
\end{array}
\right).
\end{eqnarray}

By unitary rotations, we can set $k_1=0$, i.e.
\begin{eqnarray}
\left(
\begin{array}{cccc}
&&k_2Z & m_1 \\
&&m_2 & \\
-k_2Z&-m_2&& \\
-m_1&&&
\end{array}
\right).
\end{eqnarray}

\item $N_F=4$ model C' :

%$R(\Psi_2) = R(\Psi_1)$, $R(\Psi_3) = R(\Psi_1) + 2$, $R(\Psi_4) = R(\Psi_1) + 2$, $R(\P_1)=0$.
$(R(\P_1),R(\P_2),R(\P_3),R(\P_4))=(a,a,a+2,a+2),~~a=0$.
\begin{eqnarray}
\left(
\begin{array}{cccc}
&kZ&m_3& m_1 \\
-kZ&&m_2 &m_4 \\
-m_3&-m_2&& \\
-m_1&-m_4&&
\end{array}
\right).
\end{eqnarray}

By unitary rotations, we can set $m_1=m_2=0$, i.e.
\begin{eqnarray}
\left(
\begin{array}{cccc}
&kZ&m_3& \\
-kZ&& &m_4 \\
-m_3&&& \\
&-m_4&&
\end{array}
\right).
\end{eqnarray}
\end{itemize}
All these models are in fact equivalent, and are included in $N_F=4$ model A with $k'=0$.

After all, models which satisfy all the conditions are the following three models.

$N_F=2$ model : 
\begin{eqnarray}
\left(
\begin{array}{cc}
kZ & m \\
m & \\
\end{array}
\right).
\end{eqnarray}

$N_F=4$ model B :

\begin{eqnarray}
\left(
\begin{array}{cccc}
&kZ& & m_1 \\
kZ&m_3&m_2 & \\
&m_2&& \\
m_1&&&
\end{array}
\right).
\end{eqnarray}

$N_F=4$ model C :
\begin{eqnarray}
\left(
\begin{array}{cccc}
kZ&&m_3& m_1\\
&k'Z&m_2 &m_4 \\
m_3&m_2&& \\
m_1&m_4&&
\end{array}
\right),
\end{eqnarray}
where we diagonalized the Yukawa couplings by unitary transformation.

Let us comment on CP violation.
The parameters of $N_F=2$ model and $N_F=4$ model B can be taken real without loss of generality.
Then, there is no CP violation.
However, $N_F=4$ model C has CP violation in general.
To ensure the absence of CP violation,
we assume that all parameters in the model are real.

\section{Upper bound on gaugino masses}\label{sec:3}

To avoid a cosmological problem, we want to achieve a light gravitino mass $m_{3/2} \lsim 16~\EV$.
But the lightness of the gravitino mass means smallness of $F$, which is the $F$ component of the chiral supermultiplet $Z$.
Then the gauginos become light. 

To get large gaugino masses, it is important to have a large Yukawa coupling $k$, by the following reason. 
The messenger mass spectrum depends on the combination $kF$, and the gaugino masses can be roughly written as 
\beq
m_{\rm gaugino} = \sqrt{kF}f(kF/m^2),
\eeq
where $f(kF/m^2)$ is a dimensionless function of $kF/m^2$. In gauge mediation, $f(kF/m^2)$ is bounded from above~\footnote{
Generically, the maximum value is achieved when masses of some scalar components of messenger chiral fields become almost zero.
}, $f(kF/m^2) \leq f_{\rm max}$. Then, 
the gaugino masses are also bounded from above as
\beq
m_{\rm gaugino} \leq \sqrt{kF}f_{\rm max} = f_{\rm max}\sqrt{\sqrt{3}kM_{Pl}m_{3/2}},
\eeq 
where $M_{Pl} \simeq 2.4 \times 10^{18}~\GEV$ is the reduced Planck mass, and we have assumed that the $F$ term of $Z$ is responsible for all the vacuum 
energy.
Therefore, we want large Yukawa coupling constants to achieve large enough gaugino masses.
However, we can not take arbitrary large values for the Yukawa couplings,
because the Yukawa interactions are not asymptotic free.
Then, we have to know the allowed parameter region in which the Yukawa interactions are perturbative up to the GUT scale.

Such a parameter region is determined by renormalization group (RG) equations.
The RG equations of the Yukawa couplings in each model of the previous section are given in \ref{app-B}.
By using the anomalous dimension of $Z$ and the messengers, the RG equations of the Yukawa couplings are given by
\begin{eqnarray}
\frac{dk_\chi}{dt} = \frac{k_\chi}{16\pi^2} (\gamma_Z + \gamma_{\P_\chi} + \gamma_{{\tilde \P}_\chi}) \ (\chi=d,l),
\end{eqnarray}
where $t=\log(\mu/\mu_0)$, with $\mu$ the RG scale.
$\gamma_{\P_\chi}$ and $\gamma_{{\tilde \P}_\chi}$ depend on $k_\chi^2$ for each $\chi=d,l$,
but $\gamma_Z$ is proportional to $3k_d^2 + 2k_l^2$.
This means that the RG equation of $k_l$ depends on $k_d$ through $\gamma_Z$.
Then, if we make $k_d$ smaller, we can get larger $k_l$.

First, we investigate the maximum value for the wino mass by setting $k_d=0$,
and see whether that value can exceed the CDF bound~\cite{Aaltonen:2009tp,Meade:2009qv} 
$m_{{\tilde \chi}^\pm_1} \gsim 270~\GEV$ (we will discuss on this bound later).
If $k_d \neq 0$, the wino mass bound becomes smaller than in the case $k_d = 0$.
In the case $k_d=0$, the RG equations in each model are given as follows :

$N_F=2$ model :
\begin{eqnarray}
\frac{dk_l}{dt} = \frac{k_l}{16\pi^2} \left( 4k_l^2 -3 g_2^2 - \frac{3}{5}g_1^2 \right).
\end{eqnarray}

$N_F=4$ model C:
\begin{eqnarray}
\frac{dk_l}{dt} = \frac{k_l}{16\pi^2} \left( 4k_l^2 + 2k_l'^2 -3 g_2^2 - \frac{3}{5}g_1^2 \right), \label{eq:RGlepton}\\
\frac{dk_l'}{dt} = \frac{k_l'}{16\pi^2} \left( 2k_l^2 + 4k_l'^2 -3 g_2^2 - \frac{3}{5}g_1^2 \right). \label{eq:RGlepton2}
\end{eqnarray}
The RG equation for $N_F=4$ model B is obtained by setting $k'_l=k_l$ in Eq.~(\ref{eq:RGlepton}).

By using these RG equations, we can determine the upper bounds on $k_l$ (and $k_l'$).
We require that the Yukawa coupling constants at the GUT scale are less than $4\pi$,
but our result does not strongly depend on this value as long as it is large enough.

\begin{figure}
\begin{center}
\includegraphics[scale=0.9]{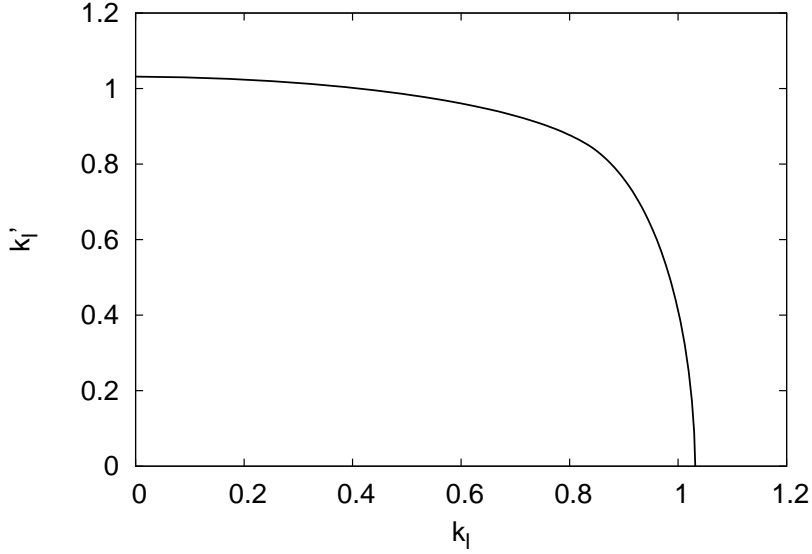}
\caption{The upper bound on the Yukawa coupling constants for $N_F=4$ model C obtained by solving the RG equations (\ref{eq:RGlepton},\ref{eq:RGlepton2}). 
The bound for $N_F=4$ model B can be obtained by setting $k'_l=k_l$.}
\label{fig:1}
\end{center}
\end{figure}

The upper bounds on the Yukawa couplings at the messenger mass scale in each model are as follows.
\begin{itemize}
\item $N_F=2$ model : 1.06
\item $N_F=4$ model C : Figure \ref{fig:1}
\end{itemize}
In these parameter regions, we calculate the wino mass.
We do not specify the mechanism generating the R-symmetry breaking vev $\vev{Z}$ (see e.g.~\cite{Dine:2006xt}),
and treat $k_l$, $m_l$ and $\la Z\ra$ as free parameters. In the calculation, we used the formulae in \ref{app-A}.
By using the upper bound on $k_l$, we can calculate the upper bound on the wino mass.

The result for the wino mass using $\alpha_2=\alpha_2(m_Z)$ is, (we will give physical pole masses later)
\begin{itemize}
\item $N_F=2$ model : $m_{{\tilde W}} \lsim \displaystyle\frac{\alpha_2}{4\pi} \sqrt{F} \times 0.24 \simeq 160\ \GEV \times \left( \frac{m_{3/2}}{16\ \EV} \right)^{1/2}$,
\item $N_F=4$ model : $m_{{\tilde W}} \lsim \displaystyle\frac{\alpha_2}{4\pi} \sqrt{F} \times 0.43 \simeq 300\ \GEV \times \left( \frac{m_{3/2}}{16\ \EV} \right)^{1/2}$,
\end{itemize}
where we have taken into account the upper bound on the gravitino mass $m_{3/2} \lsim 16\ \EV$. The upper bound is the same for both $N_F=4$ model B and C.

The actual upper bound on the wino mass (i.e. when $k_d \neq 0$) is lower than those results.
Therefore $N_F=$2 model is clearly in conflict with the CDF bound, while $N_F=4$ model is marginal. 
Let us focus on this case.

In $N_F=4$ model, we found that the gaugino masses can be maximized when $M_\chi(Z)$ is given by
\begin{eqnarray}
M_\chi(Z) = \left(
\begin{array}{cccc}
k_\chi Z&&& m_\chi  \\
&k_\chi Z& m_\chi & \\
& m_\chi&&\\
m_\chi &&&
\end{array}
\right).\label{eq:bestmodel}
\end{eqnarray}
In the rest of this section, We assume that $N_F=4$ and $M_\chi$ is of this form.
Fortunately, in this model, $m_\chi$ and $k_\chi$ can be taken real without loss of generality,
so there is no CP violation~\footnote{We thank T.~T.~Yanagida for pointing this to us.}.

Now we have six parameters, $k_d$, $k_l$, $m_d$, $m_l$, $\la Z \ra$ and $m_{3/2}$.
$k_d$ and $k_l$ are restricted by the condition that the Yukawa interactions are perturbative up to the GUT scale.
The RG equations of the model are as follows :

\begin{eqnarray}
\frac{dk_d}{dt} &=& \frac{k_d}{16\pi^2} \left( 8k_d^2 + 4k_l^2 -\frac{16}{3}g_3^2 - \frac{4}{15}g_1^2 \right),\label{eq:rgdown} \\
\frac{dk_l}{dt} &=& \frac{k_l}{16\pi^2} \left( 6k_d^2 + 6k_l^2 -3 g_2^2 - \frac{3}{5}g_1^2 \right).\label{eq:rglepton}
\end{eqnarray}

\begin{figure}
\begin{center}
\includegraphics[scale=0.9]{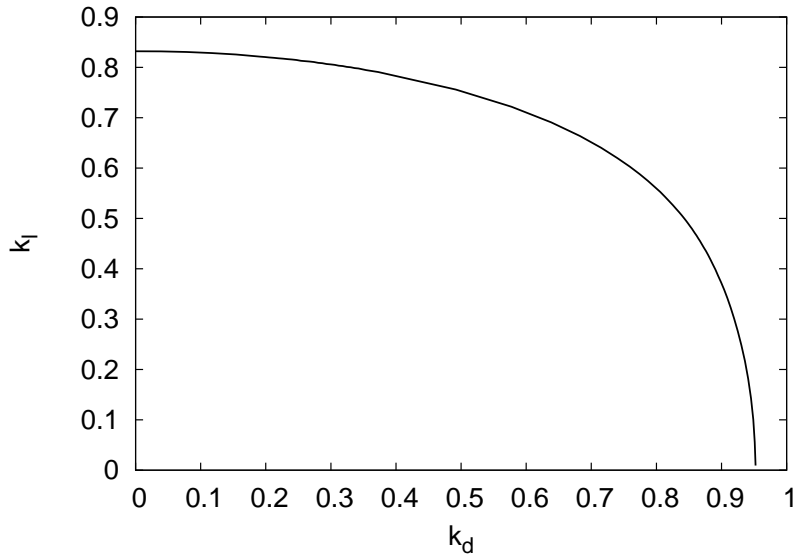}
\caption{The upper bound on the Yukawa coupling constants $k_d$ and $k_l$ obtained by solving the RG equations (\ref{eq:rgdown},\ref{eq:rglepton}).}
\label{fig:2}
\end{center}
\end{figure}

\begin{figure}
\begin{center}
\includegraphics[angle=270,scale=0.49]{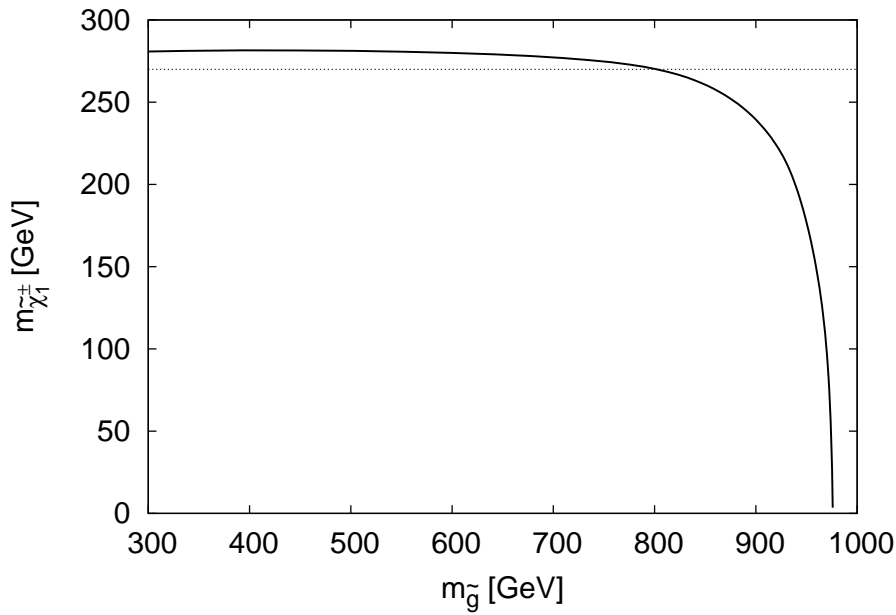}
\caption{The upper bound on the lightest chargino mass $m_{\tilde{\chi}^\pm_1}$ and the gluino mass $m_{\tilde{g}}$. 
The CDF bound $m_{\tilde{\chi}^\pm_1} \gsim 270~\GEV$ is also shown.}
\label{fig:3}
\end{center}
\end{figure}

First, we calculate the allowed parameter region of $k_d$ and $k_l$. The result is shown in Figure \ref{fig:2}.
By using it, we calculate the sparticle masses in whole the parameter space of $k_d$, $k_l$, $m_d$, $m_l$ and $\la Z \ra$ with $m_{3/2} \leq 16\ \EV$.
For the pole mass calculation, we used 
the program {\verb SOFTSUSY } 2.0.18~\cite{Allanach:2001kg}~\footnote{
We thank S. Shirai for preparing and operating the program. }. 
We assumed that ${\rm sign}(\mu)=+1$ and $\tan\beta=10$, 
where $\mu$ and $\tan\beta$ are the usual MSSM parameters (see e.g. Ref~\cite{Martin:1997ns}), but the result for the lightest chargino and the gluino masses
is nearly independent of this choice. 
We show the upper bound on the lightest chargino and the gluino masses in Figure \ref{fig:3}.

Let us discuss the experimental bound on the models discussed above. In the CDF collaboration at the Tevatron collider~\cite{Aaltonen:2009tp},
a bound on the soft masses was obtained by assuming a minimal gauge mediation model. Their parameter choice is : $N_{m}=1, M_m=2\L_m, \tan\beta=15$,
and ${\rm sign}(\mu)=+1$, where $N_m$ is the messenger number, $M_m$ the messenger mass scale, $\L_m$ the scale by which the gaugino masses are 
given by $m_{\tilde g_i}=N_m (\a_i/4\pi)\L_m$.
They obtained a bound $m_{\tilde{\chi}^0_1}>149~\GEV$ for the lightest neutralino $\tilde{\chi}^0_1$ (which is the next to lightest supersymmetric particle (NLSP)), 
and the bound can be translated into a bound on the 
mass of the lightest chargino $m_{\tilde{\chi}^\pm_1}>290~\GEV$ in that model.
In these parameters, $\tilde{\chi}^0_1$ is bino-like, 
and the next-to-lightest neutralino $\tilde{\chi}^0_2$ and the lightest chargino $\tilde{\chi}^\pm_1$ are wino-like.
Dominant production processes are $q+\bar{q} \to (\gamma/Z^0)^* \to \tilde{\chi}^+_1 \tilde{\chi}^-_1$ 
and $q+\bar{q} \to (W^\pm)^* \to \tilde{\chi}^\pm_1 \tilde{\chi}^0_2$, and they search for 
a di-photon+missing energy event which comes from the decay $\tilde{\chi}^0_1 \to \gamma \tilde{G}$ ($\tilde{G}$ is the gravitino) 
at the end of the cascade decays of produced particles.

In the models considered in this section, $\tilde{\chi}^0_1$ is the NLSP, and $\tilde{\chi}^\pm_1,\tilde{\chi}^0_2$ and $\tilde{\chi}^0_1$ are 
wino-like and bino-like respectively, except for a special parameter space discussed below. 
This is because the gauginos are somehow lighter than other sparticles due to the suppression of the gaugino masses discussed
in the Introduction. 
In Ref.~\cite{Meade:2009qv}, the CDF data was reanalyzed in the context of a general neutralino NLSP. 
They obtained a conservative bound on the lightest chargino mass $m_{{\tilde \chi}^\pm_1} \gsim 270~\GEV$ in the case $m_{\tilde B}<m_{\tilde W} \ll \mu$.
Note that the production cross section depends on the mass of $\tilde{\chi}^\pm_1$ and $\tilde{\chi}^0_2$, which is almost equal to the wino mass $m_{\tilde W}$ 
in the present model. 
The bound $m_{{\tilde \chi}^\pm_1} \gsim 270~\GEV$ exclude almost all the parameter space, as we can see from Figure~\ref{fig:3}.
Furthermore, the Tevatron may achieve a stronger bound~\cite{Meade:2009qv} 
$m_{{\tilde \chi}^\pm_1}\gsim 300~\GEV$ with $10~{\rm fb}^{-1}$ of integrated luminosity.
Even if not excluded completely at the Tevatron,
one can see from Figure~\ref{fig:3} that the gluino mass should be quite small, $m_{\tilde g} \lsim 800~\GEV$, to be discovered at the LHC. 

Let us briefly discuss other possibilities for the NLSP which is not a bino-like neutralino.
If the NLSP is not a bino-like neutralino, the branching fraction to photons is suppressed compared with the bino-like neutralino case.
One candidate for the NLSP is the stau, $\tilde{\t}_1$. We checked that the stau cannot be the NLSP for $\tan\b$ as large as $50$.
A wino-like or higgsino-like neutralino NLSP is possible in the present model (e.g. if $k_l \ll k_d$ or $k_l \gg k_d $, respectively), but then
the mass is small, $m_{\tilde{\chi}^0_1} \lsim 90~\GEV$.
This case is also excluded by the CDF bound~\cite{Meade:2009qv}. 
Also there is a region where the lightest chargino is the NLSP, but the mass is small, $m_{\tilde{\chi}^\pm_1} \lsim 90~\GEV$. Thus this case is excluded by 
LEP bounds~\cite{Amsler:2008zzb}.
Other sparticles are too heavy to be the NLSP.

Finally we comment about possible effects on the RG equations from interactions other than the messenger sector Yukawa couplings.
If $Z$ has interactions with other fields in the SUSY breaking sector, as in the IYIT model, 
$\g_Z$ presumably receives positive contribution as long as $Z$ is a gauge singlet.
This makes the messenger Yukawa couplings smaller than that obtained above, leading to a further suppression of the sparticle masses.
Introducing new singlets which couple to the messengers also make the Yukawa couplings smaller. 
On the other hand, the model Eq.~(\ref{eq:bestmodel}) has an $SU(2)$ global symmetry, which can be gauged.
The gauge interactions give negative contribution to the messenger anomalous dimensions, and
this makes the Yukawa couplings larger. Unfortunately, we found that the effect does not drastically improve the situation, 
maximally $12\%$ increase of the wino mass.
It is a model building challenge to find a mechanism which can make the Yukawa couplings larger.
One such mechanism is discussed in the next section.

\section{Direct mediation in the ISS model}\label{sec:iss}
In the previous section we have studied direct gauge mediation models under the condition that the Yukawa couplings do not blow up below the GUT scale.
Let us next consider a direct gauge mediation in the ISS model~\cite{Intriligator:2006dd}. 
In this case, Yukawa couplings are generated dynamically.
Then we can take the ``UV cutoff scale'' of the Yukawa couplings, $\L_{\rm cut}$, to be the dynamical scale of the ISS model.
The Yukawa couplings should not blow up only up to the scale $\L_{\rm cut}$.
As we lower the cutoff scale $\L_{\rm cut}$, the Yukawa couplings at the messenger scale can become larger, leading to larger gaugino masses.   
Note that this dynamical generation of the Yukawa couplings is an advantage of the ISS model over 
simply retrofitted~\cite{Dine:2006gm} O'Raifeartaigh type models.

The model we consider is based on an $SU(5)_{\rm hid}$ massive SQCD with 7 flavors of quarks and anti-quarks at high energies~\footnote{
In choosing this model, we have taken into account the Landau pole problem of the SM gauge coupling constants. See the discussion at the end of this section.
}.
After the confinement of $SU(5)_{\rm hid}$, the low-energy theory is given as follows~\cite{Seiberg:1994pq}.
The matter content of the model is; mesons $\F^I_J$ and (anti-)quarks $\v^a_I,~\tilde{\v}_a^I$. Here $I,J=1,\cdots,7$ are flavor indices
(part of which we will identify as $SU(5)_{\rm GUT}$ indices) and $a=1,2$ 
a gauge index of the dual magnetic gauge group $SU(2)_{\rm mag}$. They are decomposed as
\beq
\v^a_I=\left(
\bear{cc}
\chi^a_p &
{\P}^a_\a
\eear
\right),~~~
\tilde{\v}_a^I=\left(
\bear{c}
\tilde{\chi}_a^p \\
\tilde{\P}_a^\a
\eear
\right),~~~
\F^I_J=\left(
\bear{cc}
Y^p_q & {\P'}^p_\b \\
\tilde{\P'}^\a_q&Z^\a_\b
\eear
\right).
\eeq
Here we have decomposed the indices $I~(J)$ into $I=\{p,\a\}~(J=\{q,\b\})$, where $p=1,2$ is a flavor index and we identify $\a=1,\cdots,5$ as the $SU(5)_{\rm GUT}$ index. 

At the scale $\L_{\rm cut}$, the superpotential is given by
\beq
W&=&k\tr (\v \F\tilde{\v})-\tr(\mu^2 \F) \\
&=&k\left[\tr(\P Z\tilde{\P})+\tr(\chi\P'\tilde{\P}+\tilde{\chi}\P\tilde{\P'})+\tr(\chi Y \tilde{\chi})\right]-\tr(\mu^2 \F),
\eeq
where we have assumed that there is a single Yukawa coupling constant $k$ at the scale $\L_{\rm cut}$. 
Without loss of generality, we can assume that all parameters are real and positive.
Because of the renormalization group effect, there are several Yukawa coupling constants at low energies,
\beq
k\tr(\P Z\tilde{\P}) \to \tr\left[
\left(
\bear{cc}
\P_d & \P_l
\eear
\right)
\left(
\bear{cc}
k'_{d1}\hat{Z}_{d}+k_{d1}{\bf 1_3}\cdot Z_{d}/\sqrt{3}&k_{dl1}Z_{dl} \\
k_{dl1}Z_{ld}&k'_{l1}\hat{Z}_{l} +k_{l1}{\bf 1_2}\cdot Z_{l}/\sqrt{2} \\
\eear
\right)
\left(
\bear{c}
\tilde{\P}_d \\
\tilde{\P}_l
\eear
\right)
\right], \nonumber \\ \label{eq:issY1}
\eeq
where $\hat{Z}_{d}$ is an $SU(3)_C$ adjoint field, $\hat{Z}_{l}$ an $SU(2)_L$ adjoint field, $Z_{dl},~Z_{ld}$ bifundamental fields of $SU(3)_C\times SU(2)_L$,
and $Z_{d},~Z_{l}$ singlets. Also,
\beq
k\tr(\chi\P'\tilde{\P}+\tilde{\chi}\P\tilde{\P'}) &\to& k_{d2}\tr(\chi\P'_d\tilde{\P}_d+\tilde{\chi}\P_d\tilde{\P'}_d)+k_{l2}\tr(\chi\P'_l\tilde{\P}_l+\tilde{\chi}\P_l\tilde{\P'}_l),
\label{eq:issY2} \\
k\tr(\chi Y \tilde{\chi}) &\to& k_3\tr(\chi Y \tilde{\chi}). \label{eq:issY3}
\eeq
The RG equations for these couplings are given in \ref{app-B}. The term $\tr(\mu^2 \F)$ is, in the above notation, given by
\beq
\tr(\mu^2 \F)=\mu^2_Y\tr(Y)+\sqrt{3}\mu^2_{Zd}Z_d+\sqrt{2}\mu^2_{Zl}Z_l.
\eeq

At a SUSY breaking vacuum, $\chi,\tilde{\chi}$ develop vevs,
\beq
\vev{\chi^a_p}=\vev{\chi}\d^a_p,~~~\vev{\tilde{\chi}_a^p} = \vev{\chi}\d_a^p,~~~\vev{\chi} \equiv k_3^{-1/2}\m_Y .
\eeq
Then, we obtain a gauge mediation model
\beq
W&=&\sum_{\chi=d,l}\left[m_{\chi}\tr(\P'_{\chi}\tilde{\P}_{\chi}+\P_{\chi}\tilde{\P'}_{\chi})  \right]+k_{d1}Z_d \tr(\P_d\tilde{\P}_d)/\sqrt{3}+
k_{l1}Z_l \tr(\P_l\tilde{\P}_l)/\sqrt{2} \nonumber \\
&&~~~-\sqrt{3}\mu^2_{Zd}Z_d-\sqrt{2}\mu^2_{Zl}Z_l+\cdots,
\eeq
where $m_{l}=k_{l2}\vev{\chi}$ and $m_{d}=k_{d2}\vev{\chi}$. This is similar to the model considered in the previous section.
The vacuum energy (in global SUSY) is given by $V=3\mu^4_{Zd}+2\mu^4_{Zl}$, and the gravitino mass by $m_{3/2}=\sqrt{V}/\sqrt{3}M_{Pl}$.
We require that $m_{3/2}<16~\EV$.

There is a subtlety in choosing the values of $\mu^2_{Z\chi}~(\chi=d,l)$. If we impose that the messenger fields $\P,\tilde{\P},\P'$ and $\tilde{\P'}$ do not have
tachyonic masses, $\mu^2_{Z\chi}$ must satisfy 
\beq
\mu^2_{Z\chi} < k_{\chi 1}^{-1}m_{\chi}^2=\frac{k_{\chi 2}^2}{k_{\chi 1}k_{3}}\mu_{Y}^2.~~~~~(\chi=d,l.) \label{eq:cond1}
\eeq
On the other hand, for the above vacuum to be a global minimum of the potential at tree level, it is necessary that $\mu_{Z\chi}$ satisfy
\beq
\mu^2_{Zl} &<& \mu_{Y}^2, \label{eq:cond2-1} \\
\mu^2_{Zd}&<& \left(\frac{2k_{d1}^2+{k'}_{d1}^2}{3k_{d1}^2}\right)^{1/2}\mu_Y^2.\label{eq:cond2-2}
\eeq
These conditions can be obtained by comparing the vacuum energy at $\chi \neq 0,~\P_l=0,~\P_d=0$ (the above vacuum) 
to the vacuum energy at $\chi = 0,~\P_l \neq0,~\P_d=0$
or at $\chi = 0,~\P_l=0,~\P_d \neq 0$.
See \ref{app-D} for details.
The RG equations make Eqs.~(\ref{eq:cond2-1},\ref{eq:cond2-2}) stronger than Eq.~(\ref{eq:cond1}).
Requiring Eqs.~(\ref{eq:cond2-1},\ref{eq:cond2-2}) leads to a suppression of the gaugino masses.

\begin{figure}
\begin{center}
\includegraphics[scale=0.85]{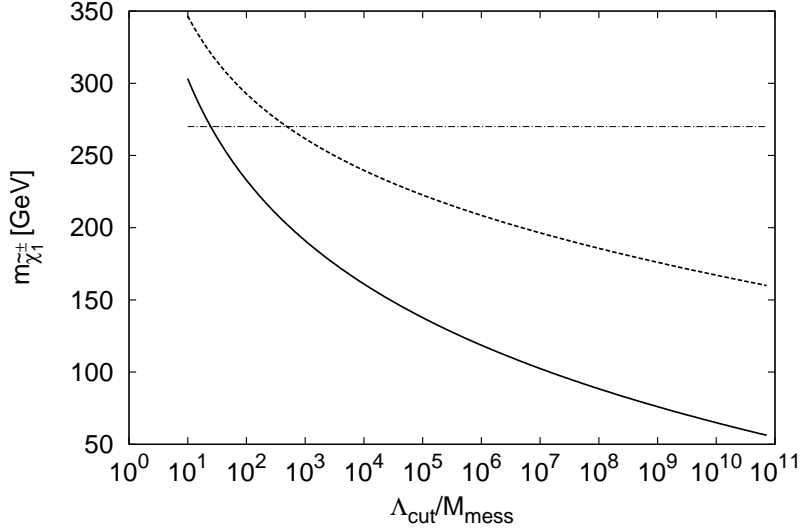}
\caption{The upper bound on the lightest chargino mass $m_{\tilde{\chi}^\pm_1}$ in the ISS model as a function of $\L_{\rm cut}/M_{\rm mess}$,
where $\L_{\rm cut}$ is the cutoff scale and $M_{\rm mess}$ the messenger mass scale.
The solid line represents the upper bound under Eqs.~(\ref{eq:cond2-1},\ref{eq:cond2-2}), while the dashed line under Eq.~(\ref{eq:cond1}). 
The CDF bound $m_{{\tilde \chi}^\pm_1} \gsim 270~\GEV$ is also shown.
The vev $\vev{Z_l}$ is chosen to maximize the lightest chargino mass.
}
\label{fig:ISSwino}
\end{center}
\end{figure}

\begin{figure}
\begin{center}
\includegraphics[scale=0.85]{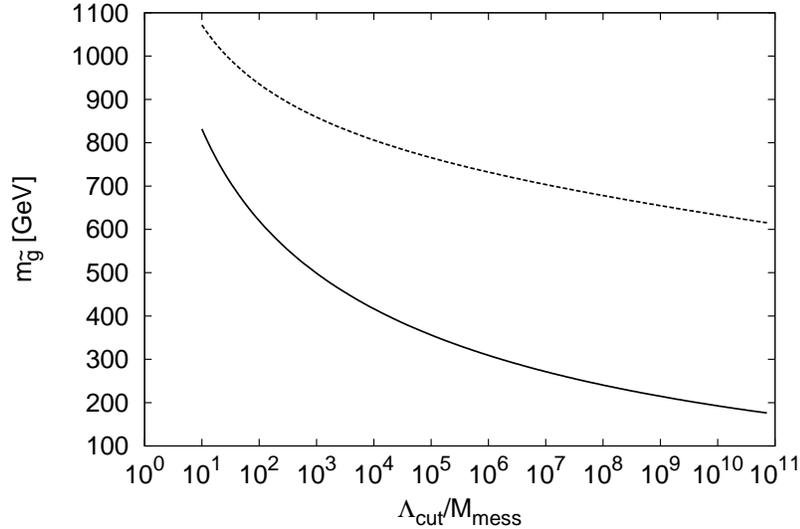}
\caption{The upper bound on the gluino mass $m_{\tilde{g}}$ in the ISS model as a function of $\L_{\rm cut}/M_{\rm mess}$.
The solid line represents the upper bound under Eqs.~(\ref{eq:cond2-1},\ref{eq:cond2-2}), while the dashed line under Eq.~(\ref{eq:cond1}). 
The vev $\vev{Z_d}$ is chosen to maximize the gluino mass.
}
\label{fig:ISSgluino}
\end{center}
\end{figure}

We calculate the maximum values of the lightest chargino and the gluino masses as a function of the cutoff scale $\L_{\rm cut}$.
We assume that all the Yukawa couplings and the magnetic gauge coupling $g_{\rm m}$ are $4\pi$ at the scale $\L_{\rm cut}$.
We do not assume model building details and simply take the R-symmetry breaking vevs $\vev{Z_d}$ and $\vev{Z_l}$ as free parameters. 
See Ref.~\cite{Kitano:2006xg} for mechanisms generating such vevs. 
The result is shown in Figure \ref{fig:ISSwino} and \ref{fig:ISSgluino}. 
We show the upper bound under either the condition Eqs.~(\ref{eq:cond2-1},\ref{eq:cond2-2}) (solid line) or Eq.~(\ref{eq:cond1}) (dashed line). However, 
for our purpose of investigating a stable SUSY breaking vacuum at perturbative level, only the solid line should be taken seriously. 
As discussed in section~\ref{sec:3}, the lightest chargino mass is bounded as
$m_{{\tilde \chi}^\pm_1} \gsim 270~\GEV$ from the CDF bound, which is possible only for a very small cutoff scale $\L_{\rm cut}/M_{\rm mess} \lsim {\cal O}(10)$.
Since the ISS model is metastable, $\L_{\rm cut}/M_{\rm mess}$ is bounded from below to suppress the vacuum tunneling rate of the model.
Detailed numerical study is required to determine how we can lower the cutoff scale $\L_{\rm cut}$. (See Ref.~\cite{Intriligator:2006dd} for a crude estimate of 
the vacuum tunneling rate.) Note also that the gluino mass is small, $m_{\tilde g} \lsim 800~\GEV$, for the cutoff scale $\L_{\rm cut}/M_{\rm mess} \gsim 10$.


Finally, let us comment on the perturbative unification of the SM gauge couplings at the GUT scale. 
Direct mediation models in the ISS model have many fields charged under the SM gauge group. 
Then, the SM gauge couplings hit Landau poles below the GUT scale
(or at least the precise perturbative unification is lost~\cite{Jones:2008ib}) 
if we consider a low scale direct mediation with $m_{3/2}<16~\EV$. Recently, it was shown that the perturbative unification can be maintained
by appropriately modifying the theory~\cite{Sato:2009yt}. In particular, a modified version of the above model
is shown to be consistent with the perturbative unification~\cite{Sato:2009yt}.
In the modified model, the low energy effective superpotential of the model is slightly different from the one described above.
It is interesting to study further in that case. See also Ref.~\cite{Abel:2008tx} for another approach to the GUT unification.

\section{Conclusions and discussion}
In this paper we have studied the parameter space of low scale direct gauge mediation models with a (perturbatively) stable SUSY breaking vacuum.
We have found that the CDF bound on the lightest chargino mass $m_{\tilde{\chi}^\pm_1} \gsim 270~\GEV$ and 
the warm dark matter bound on the gravitino mass $m_{3/2} \lsim 16~\EV$
severely restrict the allowed parameter space of the low scale direct mediation models, and almost all the parameter space is excluded. 
Even if the models somehow manage to satisfy the bound, 
the gluino mass is quite small, $m_{\tilde g} \lsim  800~\GEV$. Such a light colored particle is supposed to be discovered at the LHC experiment. 
Furthermore, in the models studied in this paper, the NLSP is a bino-like neutralino, which decays to a gravitino and a photon $\tilde{\chi}^0_1 \to \tilde{G}+\g$.
Thus there is a di-photon + missing energy signal in a SUSY event, which makes the discovery more easy.
In addition to the LHC experiment, future cosmic microwave background surveys may give a stronger upper bound on the gravitino mass~\cite{Ichikawa:2009ir}, 
$m_{3/2}\lsim 3~\EV$. Thus, the scenario studied in this paper should be completely excluded, or discovered, in near future experiments.

\section*{Acknowledgement}
We would like to thank S. Shirai and T. T. Yanagida for many useful discussions, advices, and careful reading of the manuscript.
We would also like to thank M. Endo for useful discussions.
This work was supported by World Premier International Center
Initiative (WPI Program), MEXT, Japan.  The work of KY is supported in
part by JSPS Research Fellowships for Young Scientists.

\appendix
\renewcommand{\theequation}{\Alph{section}.\arabic{equation}}
\renewcommand{\thesection}{Appendix~\Alph{section}}

\section{Runaway direction}\label{app-C}
\setcounter{equation}{0}

We will investigate the condition that a direct mediation model has no runaway direction. 
The superpotential of the model is given by
\begin{eqnarray}
W = \l\L^2 Z + {\tilde \P} (m+kZ) \P,
\end{eqnarray}
where $\det (m+kZ) = \det m \neq 0$.
We assume that the K${\rm\ddot{a}}$hler potential is canonical.
Then, the scalar potential is given by
\begin{eqnarray}
V &=& |\l\L^2 + {\tilde \P}k\P|^2 + |(m+kZ)\P|^2 + |{\tilde \P}(m+kZ)|^2.
\end{eqnarray}
We define $\k$ and ${\tilde \k}$ as
\begin{eqnarray}
\k = (m+kZ) \P, \\
{\tilde \k} =  {\tilde \P}(m+kZ).
\end{eqnarray}
Because $\det (m+kZ) = \det m \neq 0$, $(m+kZ)$ is invertible for arbitrary $Z$.
Then, by using $\k$ and ${\tilde \k}$, the scalar potential is written as 
\begin{eqnarray}
V &=& |\l\L^2 + {\tilde \k}(m+kZ)^{-1} k (m+kZ)^{-1} \k|^2 + |\k|^2 + |{\tilde \k}|^2.\label{eq1}
\end{eqnarray}
If the model has a runaway direction, in that direction each term must be arbitrarily small,
so the norm of $\k$ and ${\tilde \k}$ must be arbitrarily small.
In that direction, the first term in Eq.~(\ref{eq1}) can become small if and only if $(m+kZ)^{-1} k (m+kZ)^{-1}$ depends on $Z$.
Therefore, the condition that the model has no runaway direction leads to
\begin{eqnarray}
(m+kZ)^{-1} k (m+kZ)^{-1} &=& m^{-1}km^{-1} \nonumber \\
\Longrightarrow~~~~~~ k&=& (m+kZ)m^{-1}km^{-1}(m+kZ) \nonumber \\
&=& k + 2km^{-1}k Z + km^{-1}km^{-1}k Z^2.
\end{eqnarray}
Then, the model has no runaway direction if and only if $km^{-1}k = 0$.

\section{RG equations}\label{app-B}
\setcounter{equation}{0}

The RG equations of the Yukawa coupling constants are given as follows.
$\g_Z$ is the anomalous dimension of $Z$.

$N_F=2$ model :
\begin{eqnarray}
\frac{dk_d}{dt} &=& k_d \left( \gamma_Z + \frac{1}{16\pi^2} \left( 2k_d^2 -\frac{16}{3}g_3^2 - \frac{4}{15}g_1^2 \right) \right) \\
\frac{dk_l}{dt} &=& k_l \left( \gamma_Z + \frac{1}{16\pi^2} \left(2k_l^2 -3 g_2^2 - \frac{3}{5}g_1^2 \right) \right) \\
\gamma_Z &=& \frac{1}{16\pi^2}(3k_d^2 + 2k_l^2)
\end{eqnarray}

$N_F=4$ model C:
\begin{eqnarray}
\frac{dk_d}{dt} &=& k_d \left( \gamma_Z + \frac{1}{16\pi^2} \left( 2k_d^2 -\frac{16}{3}g_3^2 - \frac{4}{15}g_1^2 \right) \right) \\
\frac{dk_l}{dt} &=& k_l \left( \gamma_Z + \frac{1}{16\pi^2} \left( 2k_l^2 -3 g_2^2 - \frac{3}{5}g_1^2 \right) \right) \\
\frac{dk_d'}{dt} &=& k_d' \left( \gamma_Z + \frac{1}{16\pi^2} \left( 2k_d'^2 -\frac{16}{3}g_3^2 - \frac{4}{15}g_1^2 \right) \right) \\
\frac{dk_l'}{dt} &=& k_l' \left( \gamma_Z + \frac{1}{16\pi^2} \left( 2k_l'^2 -3 g_2^2 - \frac{3}{5}g_1^2 \right) \right) \\
\gamma_Z &=& \frac{1}{16\pi^2}(3k_d^2 + 2k_l^2 + 3k_d'^2 + 2k_l'^2)
\end{eqnarray}

ISS model in section~\ref{sec:iss}:
\beq
\frac{dk_{d1}}{dt}&=&\frac{k_{d1}}{16\pi^2}\left(\frac{8}{3}k_{d1}^2+\frac{16}{3}{k'}_{d1}^2+4k_{dl1}^2+4k_{d2}^2-\frac{16}{3}g_3^2 - \frac{4}{15}g_1^2 
-3g_{\rm m}^2\right)    \\
\frac{dk'_{d1}}{dt}&=&\frac{k'_{d1}}{16\pi^2}\left(\frac{2}{3}k_{d1}^2+\frac{22}{3}{k'}_{d1}^2+4k_{dl1}^2+4k_{d2}^2-\frac{34}{3}g_3^2 - \frac{4}{15}g_1^2 
-3g_{\rm m}^2\right)    \\
\frac{dk_{l1}}{dt}&=&\frac{k_{l1}}{16\pi^2}\left(3k_{l1}^2+3{k'}_{l1}^2+6k_{dl1}^2+4k_{l2}^2-3 g_2^2 - \frac{3}{5}g_1^2 -3g_{\rm m}^2\right)    \\
\frac{dk'_{l1}}{dt}&=&\frac{k'_{l1}}{16\pi^2}\left(k_{l1}^2+5{k'}_{l1}^2+6k_{dl1}^2+4k_{l2}^2-7 g_2^2 - \frac{3}{5}g_1^2 -3g_{\rm m}^2\right)    \\
\frac{dk_{dl1}}{dt}&=&\frac{k_{dl1}}{16\pi^2}\biggl(\frac{1}{3}k_{d1}^2+\frac{8}{3}{k'}_{d1}^2+\frac{1}{2}k_{l1}^2+\frac{3}{2}{k'}_{l1}^2+7k_{dl1}^2
+2k_{d2}^2+2k_{l2}^2 \nonumber \\ &&~~~~~~~~~-\frac{16}{3}g_3^2-3 g_2^2 - \frac{19}{15}g_1^2-3g_{\rm m}^2 \biggr)    \\
\frac{dk_{d2}}{dt}&=&\frac{k_{d2}}{16\pi^2}\left(\frac{1}{3}k_{d1}^2+\frac{8}{3}{k'}_{d1}^2+2k_{dl1}^2+7k_{d2}^2+2k_{l2}^2+2k_{3}^2-\frac{16}{3}g_3^2 - \frac{4}{15}g_1^2 -3g_{\rm m}^2\right)    \nonumber \\ \\
\frac{dk_{l2}}{dt}&=&\frac{k_{l2}}{16\pi^2}\left(\frac{1}{2}k_{l1}^2+\frac{3}{2}{k'}_{l1}^2+3k_{dl1}^2+3k_{d2}^2+6k_{l2}^2+2k_{3}^2-3 g_2^2 - \frac{3}{5}g_1^2 
-3g_{\rm m}^2\right)    \\
\frac{dk_{3}}{dt}&=&\frac{k_{3}}{16\pi^2}\left(6k_{d2}^2+4k_{l2}^2+6k_{3}^2 -3g_{\rm m}^2\right)    \\
\frac{dg_{\rm m}}{dt}&=&\frac{g_{\rm m}^3}{16\pi^2}
\eeq
where $g_{\rm m}$ is the gauge coupling constant of the magnetic gauge group $SU(2)_{\rm mag}$.

\section{Gaugino and sfermion mass formulae}\label{app-A}
\setcounter{equation}{0}
In this appendix, we will give formulae for the gaugino and sfermion masses~\footnote{
For general mass formulae of weakly coupled gauge mediation, see Ref.~\cite{Martin:1996zb}.
}
at the leading order of the SM gauge couplings, 
when the low energy effective superpotential is given by
\begin{eqnarray}
W = \sum_{\chi=d,l} \sum_{i,j} \left( m_{ij}^{(\chi)} {\tilde \Psi}_{i}^{(\chi)} \Psi_{j}^{(\chi)} + k_{ij}^{(\chi)} Z {\tilde \Psi}_{i}^{(\chi)} \Psi_{j}^{(\chi)} \right) .
\end{eqnarray}
After integrating out auxiliary fields, mass terms of the messengers are given by
\begin{eqnarray}
{\cal L}_{{\rm mass}} 
&=& \sum_{\chi=d,l} \Biggl\{ {\tilde \psi}^{(\chi)} m'^{(\chi)} \psi^{(\chi)} + h.c. \nonumber \\
&& +
\left(
\begin{array}{cc}
\phi^{(\chi) \dagger} & {\tilde \phi}^{(\chi)}
\end{array}
\right)
\left(
\begin{array}{cc}
m'^{(\chi) \dagger} m'^{(\chi)} & -k^{(\chi) \dagger} \langle F_Z \rangle^* \\
-k^{(\chi)} \langle F_Z \rangle & m'^{(\chi)} m'^{(\chi) \dagger}
\end{array}
\right)
\left(
\begin{array}{c}
\phi^{(\chi)} \\
{\tilde \phi}^{(\chi) \dagger}
\end{array}
\right)
\Biggr\}, \nonumber \\
\end{eqnarray}
where $m'^{(\chi)} = m^{(\chi)} + k^{(\chi)}\la Z\ra$, 
$\psi^{(\chi)}$ and ${\tilde \psi}^{(\chi)}$ denote the messenger fermions,
and $\phi^{(\chi)}$ and ${\tilde \phi}^{(\chi)}$ denote the messenger scalars.
We can diagonalize the mass matrices by unitary transformations.
There are unitary matrices $U^{(\chi)}$, $V^{(\chi)}$ and $R^{(\chi)}$ which diagonalize the mass matrices as follows :
\begin{eqnarray}
U^{(\chi) \dagger} m'^{(\chi)} V^{(\chi)} = M_{f}^{(\chi)},
\end{eqnarray}
\begin{eqnarray}
R^{(\chi) \dagger} \left(
\begin{array}{cc}
m'^{(\chi) \dagger} m'^{(\chi)} & -k^{(\chi) \dagger} \langle F_Z \rangle^* \\
-k^{(\chi)} \langle F_Z \rangle & m'^{(\chi)} m'^{(\chi) \dagger}
\end{array}
\right) R^{(\chi)}
= M_s^{(\chi)2},
\end{eqnarray}
where $M_f^{(\chi)} = \diag(m_{f,1}^{(\chi)}, m_{f,2}^{(\chi)}, \cdots)$
and $M_s^{(\chi)2} = \diag(m_{s,1}^{(\chi)2}, m_{s,2}^{(\chi)2}, \cdots)$.
We can define mass eigenstates $\phi_{d,i}^{(\chi)}$, $\psi_{d,i}^{(\chi)}$ and ${\tilde \psi}_{d,i}^{(\chi)}$ as follows :
\begin{eqnarray}
\phi_d^{(\chi)} &\equiv&
R^{(\chi) \dagger}
\left(
\begin{array}{c}
\phi^{(\chi)} \\
{\tilde \phi}^{(\chi) \dagger}
\end{array}
\right),\\
\psi_{d}^{(\chi)} &\equiv& V^{(\chi) \dagger} \psi^{(\chi)}, \\
{\tilde \psi}_{d}^{(\chi) \dagger} &\equiv& U^{(\chi) \dagger} {\tilde \psi}^{(\chi) \dagger}.
\end{eqnarray}

We define the upper half of $R^{(\chi)}$ as $A^{(\chi)}$ and the lower half of $R^{(\chi)}$ as $B^{(\chi)}$.
$A^{(\chi)}$ and $B^{(\chi)}$ are $n \times 2n$ matrices,
\begin{eqnarray}
R^{(\chi)} = \left(
\begin{array}{c}
A^{(\chi)} \\
B^{(\chi)}
\end{array}
\right). 
\end{eqnarray}
By using it, we get the formulae for the sfermion masses
\begin{eqnarray}
m^2_s &=& 2\left( \frac{\alpha_3}{4\pi} \right)^2 C_{3,s} \ m_{0}^{(d) 2}
 + 2\left( \frac{\alpha_2}{4\pi} \right)^2 C_{2,s} \ m_{0}^{(l) 2} 
 + 2\left( \frac{\alpha_1}{4\pi} \right)^2 \frac{3}{5} Y_s^2 \left( \frac{2}{5} m_{0}^{(d) 2} + \frac{3}{5} m_{0}^{(l) 2} \right) , \nonumber \label{eq:sfermionmass} \\
\end{eqnarray}
and the gaugino masses 
\begin{eqnarray}
m_{{\tilde g_1}} &=& \frac{\alpha_1}{2\pi} \left( \frac{2}{5}m_{1/2}^{(d)} + \frac{3}{5}m_{1/2}^{(l)} \right) ,\\
m_{{\tilde g_2}} &=& \frac{\alpha_2}{2\pi} \ m_{1/2}^{(l)} ,\\
m_{{\tilde g_3}} &=& \frac{\alpha_3}{2\pi} \ m_{1/2}^{(d)} .
\end{eqnarray}
In Eq.~(\ref{eq:sfermionmass}), $C_{2,s}$ and $C_{3,s}$ are the quadratic Casimir invariants, and $Y_s$ denotes the $U(1)_Y$ hypercharge.
Here $m_{0}^{(\chi) 2}$ and $m_{1/2}^{(\chi)}$ are given by
\begin{eqnarray}
m_{0}^{(\chi) 2} &\equiv& -2\sum_i m_{f,i}^{(\chi) 2} \log \ m_{f,i}^{(\chi) 2} + \sum_i m_{s,i}^{(\chi) 2} \log \ m_{s,i}^{(\chi) 2} \nonumber \\
&& - 2 \sum_{i,j} \left( \Bigl| \bigl( A^{(\chi) \dagger} V^{(\chi)} \bigr)_{ij} \Bigr|^2 + \Bigl| \bigl( B^{(\chi) \dagger} U^{(\chi)} \bigr)_{ij} \Bigr|^2 \right) m_{s,i}^2 \ {\rm Li}_2 \left( 1-\frac{m_{f,j}^{(\chi) 2}}{m_{s,i}^{(\chi) 2}} \right) \nonumber \\
&& + \frac{1}{2} \sum_{i,j} \left| \bigl( A^{(\chi) \dagger}A^{(\chi)} - B^{(\chi) \dagger} B^{(\chi)} \bigr)_{ij} \right|^2 m_{s,i}^2 \ {\rm Li}_2 \left( 1-\frac{m_{s,j}^{(\chi) 2}}{m_{s,i}^{(\chi) 2}} \right) ,\\
m_{1/2}^{(\chi)} &\equiv& \sum_{i,j} (A^{(\chi) \dagger} V^{(\chi)})_{ij} (U^{(\chi) \dagger} B^{(\chi) } )_{ji}
\frac{ m_{s,i}^{(\chi) 2} m_{f,j}^{(\chi)} }{ m_{s,i}^{(\chi) 2} - m_{f,j}^{(\chi) 2} } \log\left( \frac{ m_{s,i}^{(\chi) 2} }{ m_{f,j}^{(\chi) 2} } \right) ,
\end{eqnarray}
where ${\rm Li}_2 (x)$ is the dilog function,
\begin{eqnarray}
{\rm Li}_2(x) = \sum_{k=1}^\infty \frac{x^k}{k^2} = -\int_0^1 dt \frac{ \log(1-xt) }{t}\ \ (x\leq 1) .
\end{eqnarray}

\section{Potential minimum of the ISS model}\label{app-D}
\setcounter{equation}{0}
Due to the RG effects on the Yukawa couplings in Eqs.~(\ref{eq:issY1},\ref{eq:issY2},\ref{eq:issY3}), the potential of the ISS model is complicated.
The potential from $F$-term equations is given by
\beq
V&=&\frac{1}{3}\left| k_{d1}\tr(\tilde{\P}_d\P_d)-3\mu^2_{Zd}\right|^2+{k'}^2_{d1}\left| \tilde{\P}_d\P_d-\frac{1}{3}{\bf 1_3}\tr(\tilde{\P}_d\P_d) \right|^2 \nonumber \\
&&+\frac{1}{2}\left| k_{l1}\tr(\tilde{\P}_l\P_l)-2\mu^2_{Zl}\right|^2+{k'}^2_{l1}\left| \tilde{\P}_l\P_l-\frac{1}{2}{\bf 1_2}\tr(\tilde{\P}_l\P_l) \right|^2 \nonumber \\
&&+\left| k_{3}\tilde{\chi}\chi-\mu^2_{Y}{\bf 1_2} \right|^2 \nonumber \\
&&+k_{dl1}^2\left(\left| \tilde{\P}_d \P_l \right|^2+\left| \tilde{\P}_l \P_d \right|^2 \right) 
+k_{d2}^2\left(\left| \tilde{\P}_d \chi \right|^2+\left| \tilde{\chi} \P_d \right|^2 \right)
+k_{l2}^2\left(\left| \tilde{\P}_l \chi \right|^2+\left| \tilde{\chi} \P_l \right|^2 \right) \nonumber \\ &&+\cdots , \label{eq:isspot}
\eeq
where dots denote terms which can be set to 0 by appropriately choosing the meson vev $\vev{\F}$ (e.g. $\vev{\F}=0$).
Here $\tilde{\P}_d$ is a $3\times 2$ matrix, $\P_d$ is a $2 \times 3$ matrix, and others are $2\times 2$ matrices. 
For simplicity, let us pretend as if $k_{d2},k_{l2}$ and $k_{dl1}$ are infinitely large. Then, the last line of Eq.~(\ref{eq:isspot}) must vanish,
and we should consider the minimization of the first three lines under that condition. 

As a preparation, let us consider the minimization problem of a potential of the following form,
\beq
\hat{V}=\frac{1}{N}\left| k \tr(\tilde{\p}\p)-N\mu^2\right|^2+{k'}^2\left| \tilde{\p}\p-\frac{1}{N}{\bf 1_N}\tr(\tilde{\p}\p) \right|^2, \label{eq:prepot}
\eeq
where $\tilde{\p}$ is a $N \times n'$ matrix and $\p$ is a $n' \times N$ matrix with $N \geq n'$.
We minimize Eq.~(\ref{eq:prepot}) under the condition that ${\rm rank}(\tilde{\p}\p)= n$ with $n \leq n'$.  
By $SU(N)$ unitary transformation, we can assume without loss of generality that
\beq
\tilde{\p}\p=
\left(
\bear{cc}
A &0 \\
{*} & 0
\eear
\right), \label{eq:Aform}
\eeq
where $A$ is a $n \times n$ matrix. It is easy to see that at a potential minimum, the ${*}$ in Eq.~(\ref{eq:Aform}) is zero. Then,
\beq
\hat{V}&\geq&\frac{1}{N}\left| k \tr(A)-N\mu^2\right|^2+{k'}^2\left|A-\frac{1}{N}{\bf 1_n}\tr(A) \right|^2+{k'}^2\left|-\frac{1}{N}{\bf 1_{N-n}}\tr(A) \right|^2 \nonumber \\
&=&{k'}^2\left|A-\frac{1}{n}{\bf 1_n}\tr(A) \right|^2 \nonumber \\
&&+\left(k^2\frac{1}{N}+{k'}^2\frac{N-n}{Nn}\right)|\tr(A)|^2-2k\mu^2{\rm Re}\tr(A)+N\mu^4.
\eeq
From this form, we can see that the minimum is 
\beq
\hat{V}_{\rm min}=\frac{N(N-n){k'}^2}{nk^2+(N-n){k'}^2}\mu^4, \label{eq:issmin0} 
\eeq
at 
\beq
A=\frac{1}{n}\left(k^2\frac{1}{N}+{k'}^2\frac{N-n}{Nn}\right)^{-1}k\mu^2 {\bf 1_n} .
\eeq

By using Eq.~(\ref{eq:issmin0}), we can see that the minimum value of the potential (\ref{eq:isspot}) is given by one of the following possibilities.
\beq
V_{\rm min}=\left\{
\bear{l}
3\mu_{Zd}^4+2\mu_{Zl}^4 \\
3\mu_{Zd}^4+2\mu_{Y}^4  \\
\frac{3{k'_{d1}}^2}{2k_{d1}^2+{k'_{d1}}^2}\mu^4_{Zd}+2\mu_{Zl}^4+2\mu_{Y}^4 \\
3\mu_{Zd}^4+\frac{2{k'_{l1}}^2}{k_{l1}^2+{k'_{l1}}^2}\mu^4_{Zl} +\mu^4_Y \\
\frac{6{k'_{d1}}^2}{k_{d1}^2+2{k'_{d1}}^2}\mu^4_{Zd}+2\mu_{Zl}^4+\mu^4_Y \\
\frac{6{k'_{d1}}^2}{k_{d1}^2+2{k'_{d1}}^2}\mu^4_{Zd}+\frac{2{k'_{l1}}^2}{k_{l1}^2+{k'_{l1}}^2}\mu^4_{Zl}+2\mu_Y^4
\eear \right.
{\rm at}~{\rm rank}
(\tilde{\chi} \chi,\tilde{\P}_l\P_l,\tilde{\P}_d\P_d)=\left\{
\bear{c}
(2,0,0) \\
(0,2,0) \\
(0,0,2) \\
(0,1,1) \\
(1,0,1) \\
(1,1,0)
\eear
\right.\label{eq:minset}
\eeq
where the rank of the matrices $(\tilde{\chi} \chi,\tilde{\P}_l\P_l,\tilde{\P}_d\P_d)$ is determined by requiring that the last line of Eq.~(\ref{eq:isspot})
should vanish. By solving the RG equations presented in \ref{app-B} numerically, we checked that $k_{d1}<k'_{d1}$ and $k_{l1}<k'_{l}$. 
Then, requiring that the first of Eq.~(\ref{eq:minset}) is the smallest, we obtain 
\beq
\mu^2_{Zl} &<& \mu_{Y}^2, \label{eq:condD-1} \\
\mu^2_{Zd}&<& \left(\frac{2k_{d1}^2+{k'}_{d1}^2}{3k_{d1}^2}\right)^{1/2}\mu_Y^2.\label{eq:condD-2}
\eeq
Under these conditions, the vacuum with ${\rm rank}(\tilde{\chi} \chi,\tilde{\P}_l\P_l,\tilde{\P}_d\P_d)=(2,0,0)$ is stable.

So far we have assumed that $k_{d2},k_{l2}$ and $k_{dl1}$ are infinitely large. It is obvious that Eqs.~(\ref{eq:condD-1},\ref{eq:condD-2})
are necessary for the above vacuum to be stable even without that assumption. But it is slightly complicated to determine whether these are also sufficient 
or not. After a little lengthy calculations, we checked that if the conditions
\beq
\mu^2_{Zd}&<&\mu^2_Y \min\left\{\frac{k_{d2}^2}{k_3k_{d1}},\frac{k_{dl1}^2}{k_{l1}k_{d1}} \right\},\\
k_{3}k_{l1}&<&k_{l2}^2,
\eeq
are satisfied, Eqs.~(\ref{eq:condD-1},\ref{eq:condD-2}) are also sufficient to ensure the stability of the above vacuum.
The numerical solutions to the RG equations satisfy these conditions.

\end{document}